\documentclass[12pt]{article}

\usepackage{graphicx}
\usepackage{subfigure}
\usepackage{algpseudocode}
\usepackage{amsfonts}
\usepackage{amsmath, amsthm, amssymb}
\usepackage{dsfont}
\usepackage[utf8x]{inputenc}
\usepackage{color}
\usepackage{geometry}
\usepackage{hyperref}
\usepackage{ifthen}
\usepackage{algpseudocode}
\usepackage{float}

\usepackage{setspace}
\title{Regular graphs maximize the variability of random neural networks}
\author{Gilles Wainrib\footnote{Ecole Normale Sup\'erieure, D\'epartement d'Informatique, \'equipe DATA, Paris, France.} \and Mathieu Galtier \footnote{European Institute for Theoretical Neuroscience, Paris, France.}}

\begin{document}

\maketitle
%\tableofcontents

 \begin{abstract}
	\textit{
In this work we study the dynamics of systems composed of numerous interacting elements interconnected through a random weighted directed graph, such as models of random neural networks. We develop an original theoretical approach based on a combination of a classical mean-field theory originally developed in the context of dynamical spin-glass models, and the heterogeneous mean-field theory developed to study epidemic propagation on graphs. Our main result is that, surprisingly, increasing the variance of the in-degree distribution does not result in a more variable dynamical behavior, but on the contrary that the most variable behaviors are obtained in the regular graph setting. We further study how the dynamical complexity of the attractors is influenced by the statistical properties of the in-degree distribution.}
 \end{abstract}

\section{Introduction}

The modeling of systems involving many dynamical units connected by a complex network of interactions is a topic of increasing interest in many scientific domains, such as neuroscience \cite{sompolinsky1988chaos}, genetics \cite{kauffman1969metabolic,bower2001computational},  epidemiology \cite{pastor2001epidemic}, artificial intelligence \cite{jaeger2004harnessing} or social sciences \cite{castellano2009statistical} (see also \cite{barrat2008dynamical} for a comprehensive overview). In these systems, depending on the nature of the interactions one can observe various complex phenomena such as the order-disorder phase transition in spin-glasses, the propagation of an epidemic or the emergence of complex dynamics in neural networks models. In general, understanding the emerging properties of such systems relies on the combination of three main factors : (i) the structure of the connectivity network, (ii) the way units interact with each other and (iii) the dynamical properties of single units. One of the main theoretical challenges in this line of research is to unravel the subtle links between these three factors.

A particular question of interest concerns the relationship between the degree distribution of a weighted graph and the properties of a system of interacting non-linear dynamical units on such a graph. For example, this topic has been studied recently in the context of synchronization phenomenon, showing the importance of the degree distribution for characterizing the transition to the synchronous state \cite{roxin2011role, sonnenschein2013excitable, sonnenschein2012onset}.

In this paper, we consider a discrete-time non-linear neural network model which is used in various artificial intelligence applications based on artificial neural networks, and corresponds to the class of firing-rate models commonly used in computational neuroscience. Despite its apparent simplicity, this model provides an interesting theoretical framework to study analytically the influence of the structure of the connectivity on the dynamical properties of the system. The study of this class of models on fully connected random networks \cite{sompolinsky1988chaos} through the mean-field approach reveals a phase-transition between stable steady states and chaotic dynamics in the limit of infinite networks. However, the impact of an underlying graph structure on such random weights models has not been studied so far. In this situation, the classical mean-field approach needs to be augmented with ideas coming from the field of theoretical epidemiology, where several authors have suggested to group nodes by their in-degrees \cite{pastor2001epidemic} giving rise to the emerging method called heterogeneous mean-field theory. In this article, we develop upon this approach constructing an original method to derive consistent equations governing the variability of the network nodes. A consequence of our theoretical analysis is that regular graphs are shown to enhance the variability of such dynamical systems, while making the system more homogeneous in terms of network topology. We also show how to quantify the complexity of the system's attractors, by explicitly computing the largest Lyapunov exponents of the system.

In Section \ref{sec: HMF}, we introduce formally our model and present an analytical study of a classical recurrent neural network model. In section \ref{sec: simus}, we provide numerical simulations verifying and extending the results to other systems, such as networks of interacting FitzHugh-Nagumo neurons.

\section{Heterogeneous mean-field theory} \label{sec: HMF}
Here we develop an original analytical approach, which we dub Heterogeneous Mean-Field (HMF) theory, for the study of randomly connected recurrent neural networks.
\subsection{Model}\label{sec: model}
Consider a directed graph $G=(v,E)$ with $n$ nodes, and an in-degree distribution $P_n(k)$. We denote $\alpha=k/n \in [0,1]$ the rescaled in-degree and assumes that in the large $n$ limit, $P_n(n\alpha) \to p(\alpha)$, called the rescaled in-degree distribution (the typical in-degree of a node is assumed to be of order $n$). We further make the assumption of no correlation between in- and out-degree. To our knowledge the question of assortative mixing properties of anatomical neuronal networks has not been fully elucidated so far \cite{bullmore2009complex}, and remains a question of current research.  On each edge $e=j\to i$, one assigns a weight $J_{ij}$, which are independent centered random variables with finite variances, which may depend on the in-degree $\alpha$ of node $i$: $Var[J_{ij}] = \sigma^2_{\alpha}/n$. We consider the following dynamical system on this weighted di-graph:
\begin{equation}
	x_i(t+1) = S\left(\sum_{j\to i} J_{ij} x_j(t)\right)
	\label{eq:NN}
\end{equation}
where $S(.)$ is an odd sigmoid function with $S(0)=0$, $S'(0)=1$,$S'(x)>0$ and $x S''(x)\leq 0$ (for instance $S(x)=\tanh(x)$ in a popular choice in the literature). This model can be seen as a dynamical version of a zero-temperature spin-glass model \cite{mezard1987spin} and has been studied in the context of neural networks in \cite{sompolinsky1988chaos,cessac1994mean,wainrib2013optimal,wainrib2013topological} in the case where the graph is fully connected, i.e. $p=\delta_1$.

\begin{figure}[ht!]
	\centering
\subfigure[$c=0.5$ and $Var(\alpha)=0.0089$]{\includegraphics[width = 8cm]{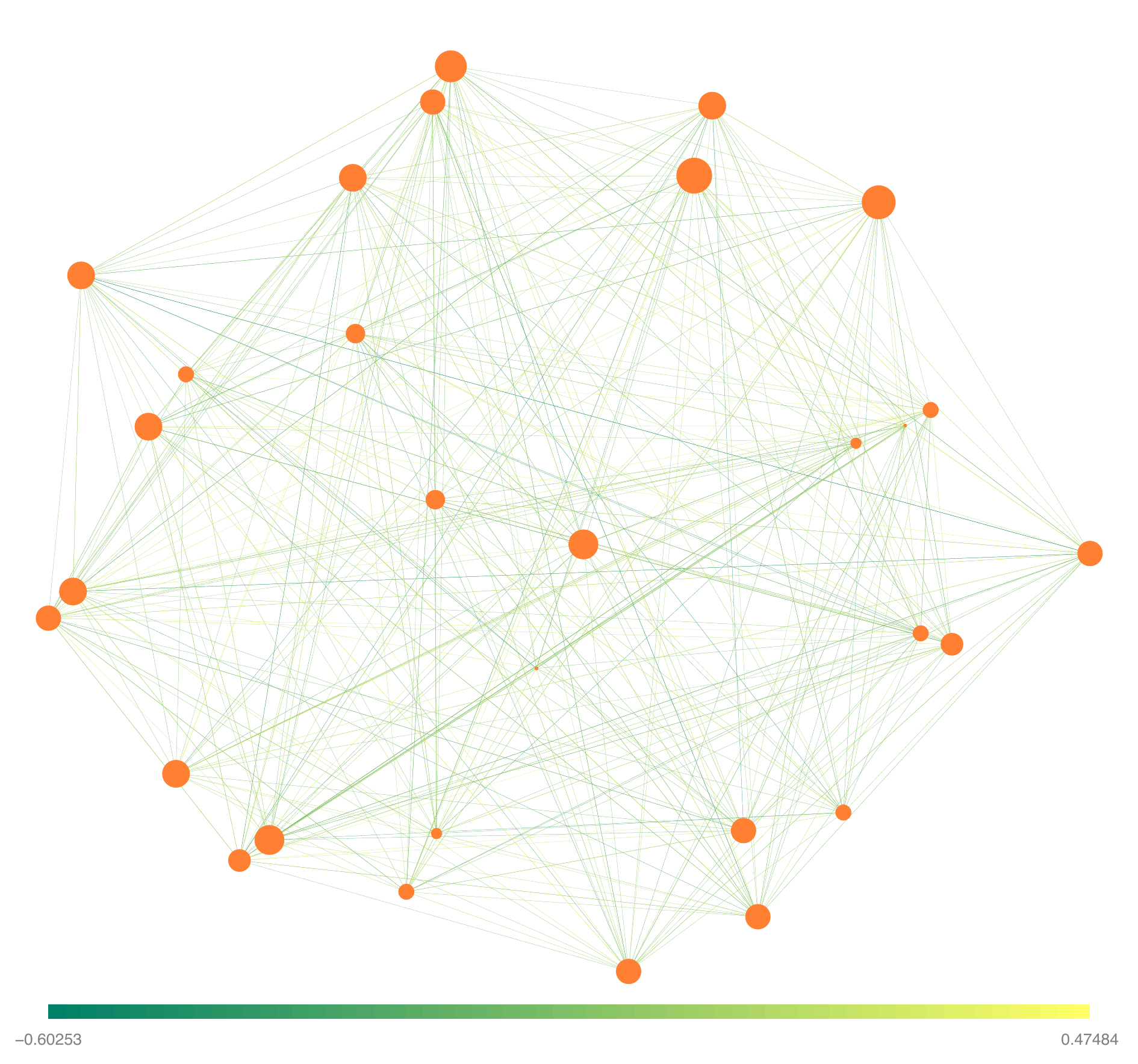}} 
\subfigure[$c=0.25$ and $Var(\alpha)=0.0639$]{\includegraphics[width = 8cm]{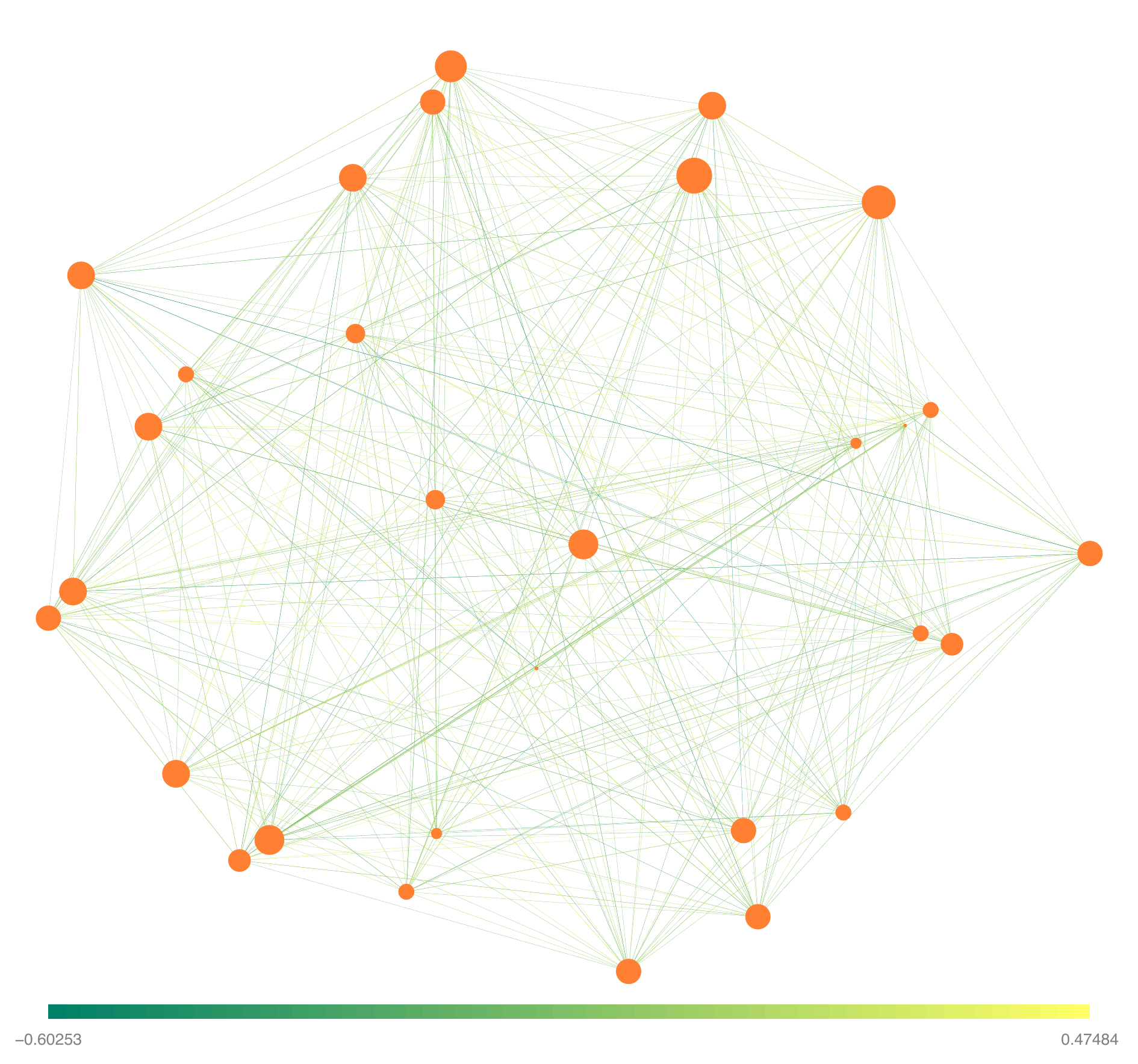}}\\
\subfigure[$c=0.1$ and $Var(\alpha)=0.1634$]{\includegraphics[width = 8cm]{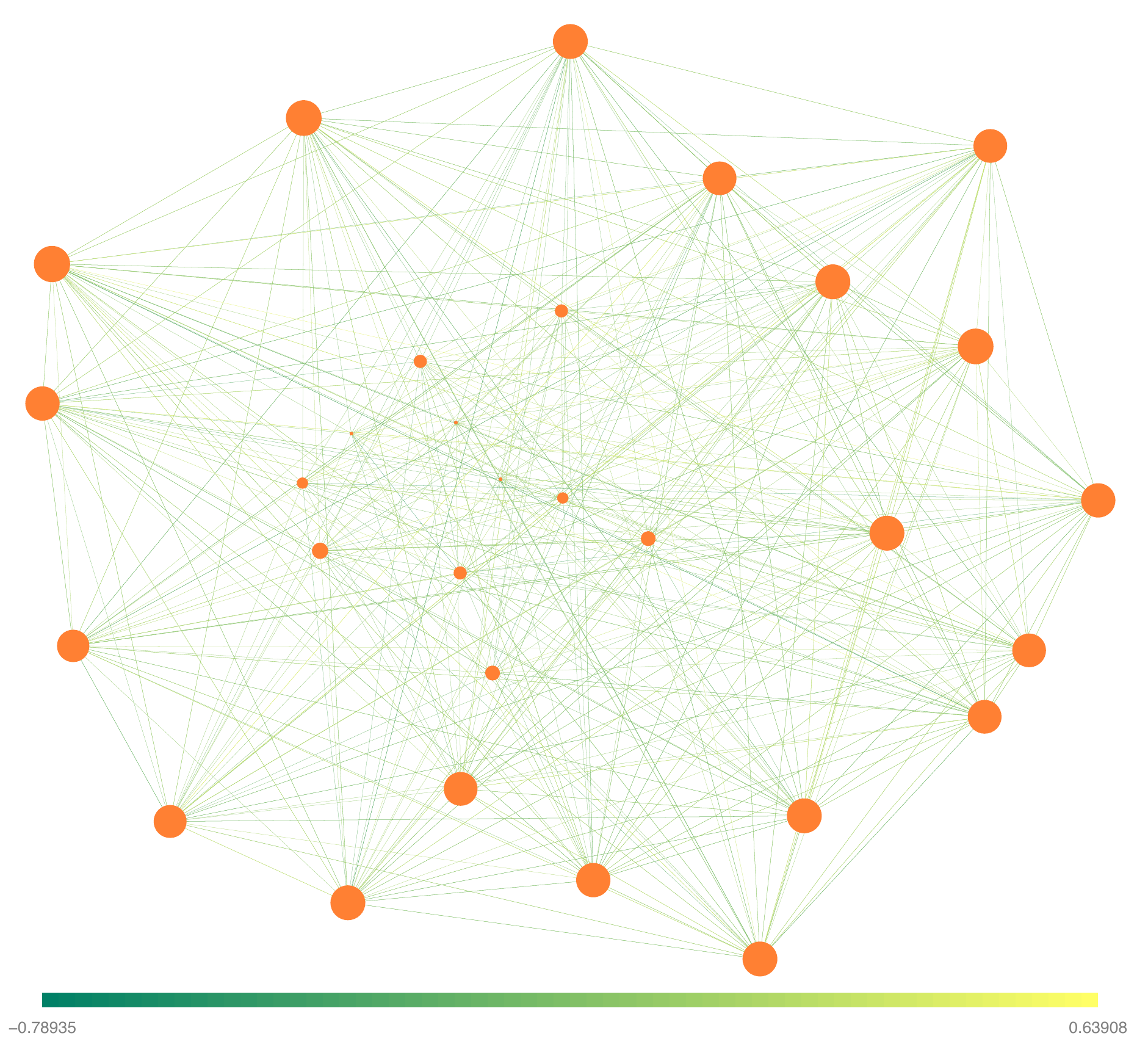}}
\subfigure[$c=0.01$ and $Var(\alpha)=0.2401$]{\includegraphics[width = 8cm]{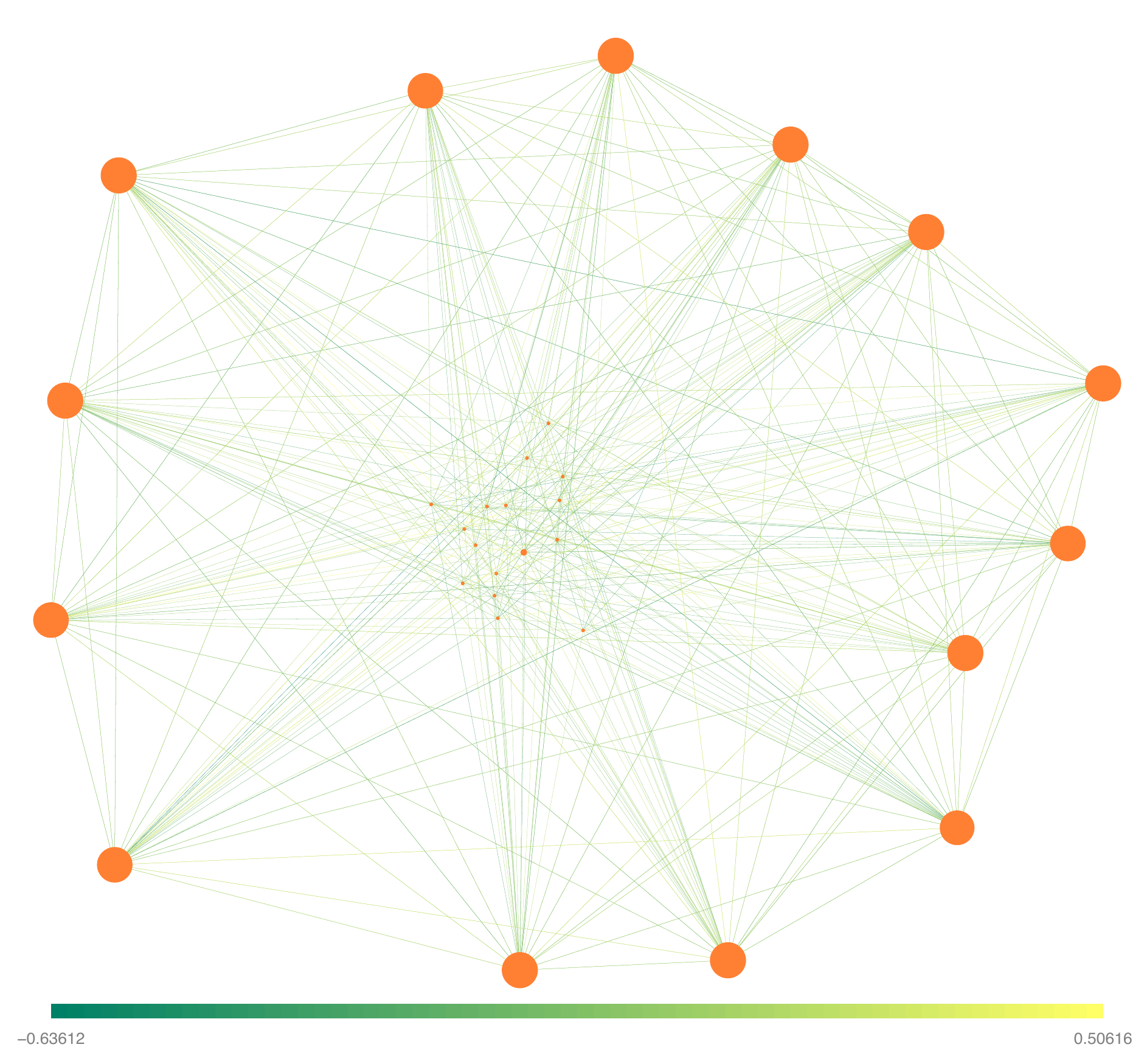}} 
\caption{Sample networks of $n=30$ neurons with various in-degree distributions, drawn from $p=(\delta_c + \delta_{1-c}) /2$ for different values of $c$. The size of each vertex is proportional to the number of incoming connections (in-degree), which is on average equal to $n/2=15$ by definition of $p$. Edge color represents random weights $J_{ij}$.}
\label{fig:net}
\end{figure}

\subsection{General HMF equations}
Here, we study the previous system deriving a self-consistent equation characterizing the variance of the nodes variables $x_i(t)$, showing how a combination of the degree distribution $p$ and the variance profile $\sigma^2_{\alpha}$ controls the transition from a stable null equilibrium to a disordered state. 

In the case of a general weighted graph, the overall strategy we introduce is to combine the classical mean-field approach \cite{sompolinsky1981dynamic,sompolinsky1988chaos,cessac1994mean} with the idea of partitioning the nodes according to their rescaled degree $\alpha$, at the heart of the heterogeneous mean-field (HMF) theory developed in the field of epidemiology \cite{pastor2001epidemic,vespignani2011modelling}. Suppose $i$ is a node with rescaled degree $\alpha$ and denote $\gamma_{\alpha}^2(t)$ the variance of $x_i(t)$ and $a_i=\sum_{j\to i} J_{ij} x_j(t)$. This sum contains only $n\alpha$ terms, which come from nodes with various degree, and not only from nodes of degree $\alpha$. The main idea of HMF is that these terms effectively sample the overall behavior of the system. Then the key step in the classical mean-field approach is to assume the independence between the $x_i$'s and $J$ (mean-field assumption). This has been rigorously justified in \cite{arous1995large,moynot2002large,cabana2013large} using large deviation techniques when adding an arbitrary small white-noise term on \eqref{eq:NN}, but remains an open problem in the zero-noise case. Under the mean-field assumption, one deduces that $a_i$ behaves as a centered Gaussian variable with variance $\alpha \sigma^2_{\alpha} \gamma^2(t)$ where 
\begin{equation}
	\gamma^2(t)=\int_0^1 p(\alpha) \gamma^2_{\alpha}(t)d\alpha
\end{equation}
From the iteration equation $x_i(t+1)=S(a_i(t))$, we further deduce:
\begin{equation}
	\gamma^2_{\alpha}(t+1)=F(\alpha\sigma^2_{\alpha}\gamma^2(t))
\end{equation}
with \begin{equation}
	F(z^2) = (2\pi)^{-1/2}\int_R S^2(zx)e^{-x^2/2}dx
\end{equation}
Combining the above equations, we obtain:
\begin{equation}
	\gamma^2(t+1)=\int_0^1 p(\alpha)F(\alpha\sigma^2_{\alpha}\gamma^2(t)))d\alpha := \bar{F}(\gamma^2(t))
	\label{eq:rec}
\end{equation}
An interesting case occurs when choosing a specific type of non-linear activation function $S(.)$. Indeed, when $S(x)=\mbox{erf}\left(\frac{\sqrt{pi}}{2}x\right)$, it is possible to compute explicitly the Gaussian integral defining function $F$, yielding:
\begin{equation}
	F(z^2)=\frac{2}{\pi}\arcsin\left(\frac{\pi z^2}{2+\pi z^2}\right)
\end{equation}
Therefore, if one considers for instance a distribution $p$ which is the sum of two Dirac deltas $p=(\delta_c + \delta_{1-c}) /2$, where $c \in [0,1]$, one obtains an explicit update equation for the variance:
\begin{equation}
	\label{eq:gamma}
		\gamma^2(t+1)= \frac{1}{\pi}\arcsin\left(\frac{\pi c\sigma^2 \gamma^2(t)}{2+\pi c\sigma^2 \gamma^2(t)}\right) + \frac{1}{\pi}\arcsin\left(\frac{\pi (1-c)\sigma^2 \gamma^2(t)}{2+\pi (1-c)\sigma^2 \gamma^2(t)}\right)
\end{equation}
further assuming a constant variance profile $\sigma_{\alpha}^2=\sigma^2$.
\subsection{Phase transition}
Therefore, to understand the order-disorder phase transition in this system, the first step is to study the dynamical system $\gamma^2(t+1)=\bar{F}(\gamma^2(t))$. Due to the properties of the sigmoid function $S(.)$, the function $F$ is increasing, concave and satisfies $F(0)=0$ and $F'(0)=1$. We then deduce that $\bar{F}$ is also increasing, concave and satisfies $\bar{F}(0)=0$. Therefore, beyond the trivial equilibrium $\gamma^2=0$, the existence of another non-trivial equilibrium for \eqref{eq:rec} will depend on the value of the slope at zero:
 \begin{equation}
	\bar{F}'(0)=\int_0^1 p(\alpha)\alpha\sigma^2_{\alpha}d\alpha := \mu
\end{equation}
Therefore, $\gamma^2(t)$ converges to $\gamma^2_{\infty}=0$ if $\mu < 1$, and $\gamma^2(t)$ converges to a limit value $\gamma^2_{\infty}>0$ if $\mu > 1$. 
For instance, in the classical case of a complete graph with homogeneous variances, $p(\alpha)=\delta_1$ and $\sigma^2_{\alpha}=\sigma^2$, so the critical condition becomes $\sigma=1$ as expected. 
In the case of a regular graph with rescaled degree $\alpha_0$ (each node has $\lceil \alpha_0 n \rceil$ incoming edges) with homogeneous variances, $p(\alpha)=\delta_{\alpha_0}$ and $\sigma^2_{\alpha_0}=\sigma^2$, so the critical variance parameter becomes $\sigma=\alpha_0^{-1/2}$. 
In the case of a general degree distribution with homogeneous variances $\sigma^2_{\alpha}=\sigma^2$ for all $\alpha \in [0,1]$, one obtains that the critical value of $\sigma$ is given by $\langle \alpha \rangle^{-1/2}$ where $\langle \alpha \rangle$ is the mean rescaled degree. Finally, among all the possible choices of $\sigma_{\alpha}$, the case $\sigma^2_{\alpha}=\sigma^2/\alpha$ is particularly interesting since the critical value is always $\sigma=1$, whatever the choice of $p(\alpha)$, as one would have expected. This choice might correspond to the concept of synaptic scaling \cite{turrigiano2000hebb}, ensuring that the overall input coming to a given unit has a typical strength independent of its in-degree. 

The critical parameter $\mu$ can be seen as a weighted average of the rescaled degrees, and in the case of a homogeneous variance profile is precisely proportional to the averaged rescaled degree $\sigma^2 \langle \alpha \rangle$. In this case, it is natural to investigate the impact of the rescaled degree higher moments, such as its variance $Var(\alpha)$, on the value of the fixed point $\gamma^2_{\infty}$ characterizing the disordered state. It is not possible in general to obtain a closed form expression of $\gamma^2_{\infty}$, however close to the transition, $\bar{F}$ can be approximated by $\bar{F}(x)=x\bar{F}'(0)+\frac{x^2}{2}\bar{F}''(0)+O(x^3)$ with $\bar{F}'(0)$ given above, and $\bar{F}''(0)=F''(0)\int_0^1p(\alpha)\alpha^2\sigma^4_{\alpha}d\alpha <0$ since $F''(0)<0$. Restricting the analysis to the homogeneous variances case, that is $\sigma^2_{\alpha}=\sigma^2$, we introduce a small parameter $\epsilon=\sigma^2\langle \alpha \rangle-1$ and obtain:
\begin{equation}
	x = x\sigma^2\langle \alpha \rangle + \frac{F''(0)\sigma^4}{2}(Var(\alpha)+\langle \alpha \rangle^2)x^2+O(x^3)
\end{equation}
\begin{equation}
	\gamma^2_{\infty}\sim \frac{2\epsilon}{-F''(0)\sigma^4(Var(\alpha)+\langle \alpha \rangle^2)}=a_1\epsilon
\end{equation}
Therefore, for a given mean-degree $\langle \alpha \rangle$, this formula implies that increasing the variance $Var(\alpha)$ will decrease the variability $\gamma^2_{\infty}$ of the neural activity variables, implying that regular graphs maximize $\gamma^2_{\infty}$ when the mean-degree is kept fixed. 

In fact, it is also possible to compute the $O(\epsilon^2)$ term in the expansion of $\gamma_{\infty}^2=a_1 \epsilon + a_2\epsilon^2 + O(\epsilon^3)$ and one finds:
\begin{equation}
	a_2=\frac{2F'''(0)}{3\sigma^6F''(0)^3}\frac{\langle \alpha^3 \rangle}{\langle \alpha^2 \rangle^3}
\end{equation}
\subsection{Lyapunov exponent}
To further characterize the non-trivial attractors in the disordered phase, it is of interest to estimate the maximal Lyapunov exponent $\lambda$, as in \cite{PhysRevLett.69.3717},  which is below 1 in the case of a steady state, equal to 1 in the case of a limit cycle and larger than 1 in the case of a chaotic attractor. To this end, we consider two solutions of (1) starting from two different initial conditions such that $a_i(0)-a'_i(0) \sim N(0,\delta(0)^2)$. Denoting $\Delta_i(t)=a_i(t)-a'_i(t)$, we have:
\begin{eqnarray*}
	&&\Delta_i(t+1)=\sum_{j\to i}J_{ij}(S(a_j(t))-S(a'_j(t)))\\
	&=& \sum_{j\to i}J_{ij}S'\left((a_j(t)+a'_j(t))/2\right)\Delta_j(t)+O(||\Delta(t)||^2)
\end{eqnarray*}
Therefore, if node $i$ has a rescaled in-degree $\alpha$, one obtains the following relationship on the variances:
\begin{equation}
	\delta_{\alpha}^2(t+1) = \alpha \sigma^2_{\alpha} \Phi(\alpha\sigma^2_{\alpha}\gamma^2(t)) \delta^2(t)+O(\delta^3(t))
\end{equation}
with 
\begin{equation}
	\Phi(z^2) := (2\pi)^{-1/2}\int S'^2(z x)e^{-x^2/2}dx
\end{equation}
Integrating with respect to the degree distribution, we obtain:
\begin{equation}
	\delta^2(t+1) = \bar{\Phi}(\gamma^2(t)) \delta^2(t)+O(\delta^3(t))
\end{equation}
with \begin{equation}
	\bar{\Phi}(\gamma^2(t)):=\int_0^1p(\alpha)\alpha \sigma^2_{\alpha} \Phi(\alpha\sigma^2_{\alpha}\gamma^2(t))d\alpha
	\label{eq:phi}
\end{equation}
On the attractor, $\gamma^2(t)=\gamma^2_{\infty}$ so we deduce that 
\begin{equation}
	\lambda:=\lim_{t\to \infty,\delta(0)\to 0} \left(\frac{\delta^2(t)}{\delta^2(0)}\right)^{1/t} = \bar{\Phi}(\gamma^2_{\infty})
\end{equation}
From this formula, one deduce, as expected, that in the subcritical regime $\lambda=\mu<1$ because $\gamma^2_{\infty}=0$ and $\Phi(0)=1$. In the supercritical regime, it is natural to wonder whether $\lambda$ will also be a decreasing function of $Var(\alpha)$. The answer is not straightforward because the distribution $p$ appears in $\lambda$ both explicitly in formula \eqref{eq:phi} and implicitly through the dependence of $\gamma_{\infty}^2$. Denoting $\beta=2/(-F''(0)\sigma^4\langle \alpha^2 \rangle)$ and $\epsilon=\sigma^2\langle \alpha \rangle-1$, we know that $\gamma^2_{\infty}\sim \beta\epsilon$ when $\epsilon \ll1$. Therefore, from a Taylor expansion of $\Phi$ around zero, we obtain:
\begin{eqnarray*}
	\bar{\Phi}(\gamma_{\infty}^2)&=&1+\epsilon(1+\langle \alpha^2 \rangle\sigma^2 \beta \Phi'(0)) \\&+&\epsilon^2(\langle \alpha^3 \rangle\sigma^4\beta^2/2) + O(\epsilon^3)
\end{eqnarray*}
When substituting $\beta$ with its expression, one discovers that $\langle \alpha^2 \rangle$ disappears in the $O(\epsilon)$ term, and appears next in the $O(\epsilon^2)$ term:
\begin{eqnarray*}
	&&\bar{\Phi}(\gamma_{\infty}^2)=1+\epsilon(1-2\frac{\Phi'(0)}{F''(0)\sigma^2}) \\&+&\epsilon^2\frac{\langle \alpha^3 \rangle}{\langle \alpha^2 \rangle^2}\frac{1}{\sigma^4}\left(\frac{2\Phi''(0)}{ F''(0)^2}+\frac{2F'''(0)\Phi'(0)}{3F''(0)^3}\right) + O(\epsilon^3)
\end{eqnarray*}
The first consequence of this formula is that, at first order close to the transition, the dynamical complexity of the trajectories as measured by the Lyapunov exponent, does not depend on the higher order moments of the degree distribution. This means that $Var(\alpha)$ controls the variability of the variables $x_i$ without modifying complexity of the chaotic attractor. This is true only at first order in $\epsilon$, and the dependence of the Lyapunov exponent is expressed through a more complicated parameter $\nu:=\langle \alpha^3 \rangle/\langle \alpha^2 \rangle^2$, which accounts for the higher moments of the rescaled degree distribution, beyond the variance.

\section{Numerical simulations}\label{sec: simus}

\subsection{Validation of the theory}
We have tested the theoretical predictions of the mean-field theory by numerical simulations, in particular the claim that regular graphs should maximize the variability of the nodes variables. To this end, we construct random weighted graphs with a prescribed degree distribution $p_c=\frac{1}{2}(\delta_c + \delta_{1-c})$, by selecting half of the nodes to have an in-degree $cn$ and the other half with an in-degree $(1-c)n$. We have chosen this distribution because it combines simplicity with the ability to manipulate the variance of the in-degrees without changing the average. Indeed, in this setting, the average rescaled degree $\langle \alpha \rangle$ is kept constant at $1/2$, while the variance $Var(\alpha)$ depends on $c$ as $Var(\alpha)=(c-1/2)^2$. To measure the variability of the node variables, we compute the temporal average of the instantaneous variances of $x_i(t)$:
\begin{equation}
	\hat{\gamma}^2=\frac{1}{nT}\sum_{t=1}^T \sum_{i=1}^n (x_i(t)-\bar{x}(t))^2
\end{equation}
where $\bar{x}(t)=\frac{1}{n}\sum_{i=1}^n x_i(t)$. Results are summarized in Figure 1, where the estimated variance $\hat{\gamma}^2$ is displayed against the variance of the rescaled degree $Var(\alpha)$, showing that the variability of the node variables is maximal when the graph is regular. The fairly good agreement between the numerical simulations and the values predicted by the theory (computed with equation \eqref{eq:gamma}) hence provides a strong support for the theoretical calculations obtained in Section 2, even if the simulation was done away from the order-disorder transition ($\sigma = 2$).
\begin{figure}
	\label{fig:1}
	\begin{center}
	\includegraphics[width=14cm]{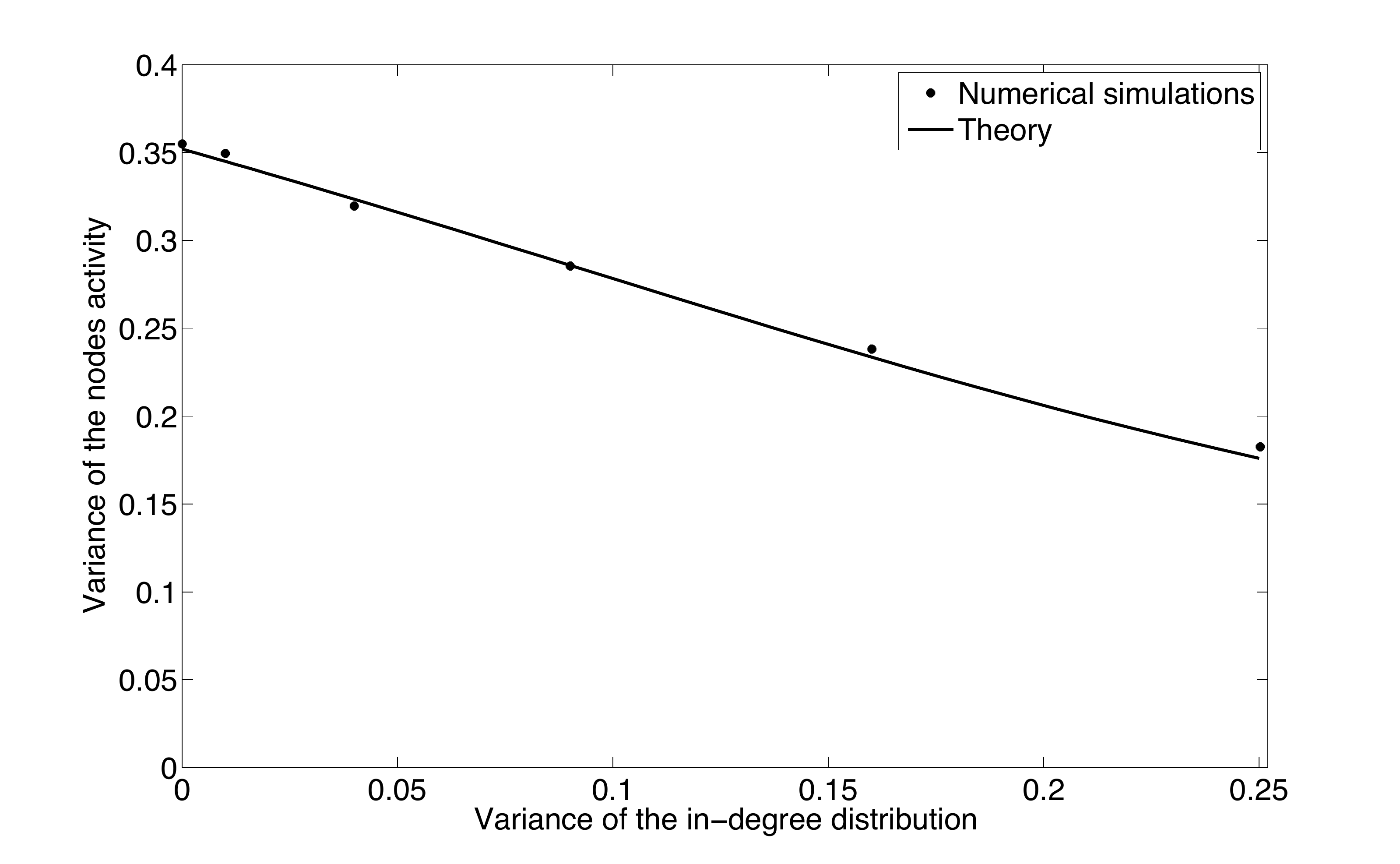}
	\caption{Theoretical variance $\gamma^2_{\infty}$ and numerically estimated variance $\hat{\gamma}^2$ as a function of the in-degree variance $Var(\alpha)$. Parameters $n=1000$, $\sigma=2$. The family of in-degree distribution chosen here is $p=(\delta_c + \delta_{1-c})/2$, and the parameter $c$ is varied between $0$ and $0.5$ to generate networks with different variances $\mbox{Var}(\alpha)$ of the in-degree distribution.}
\end{center}
\end{figure}

\subsection{Limitations and extensions}
The results obtained so far are only focused on a specific, yet important for various applications, neural network model, and it is legitimate to investigate whether the results extend to other models. In particular, it is natural to check if the qualitative phenomenon of maximal variability for regular graphs presented above still holds when the dynamical model of the nodes is different.

\begin{figure}
	\label{fig:2}
	\begin{center}
	\includegraphics[width=14cm]{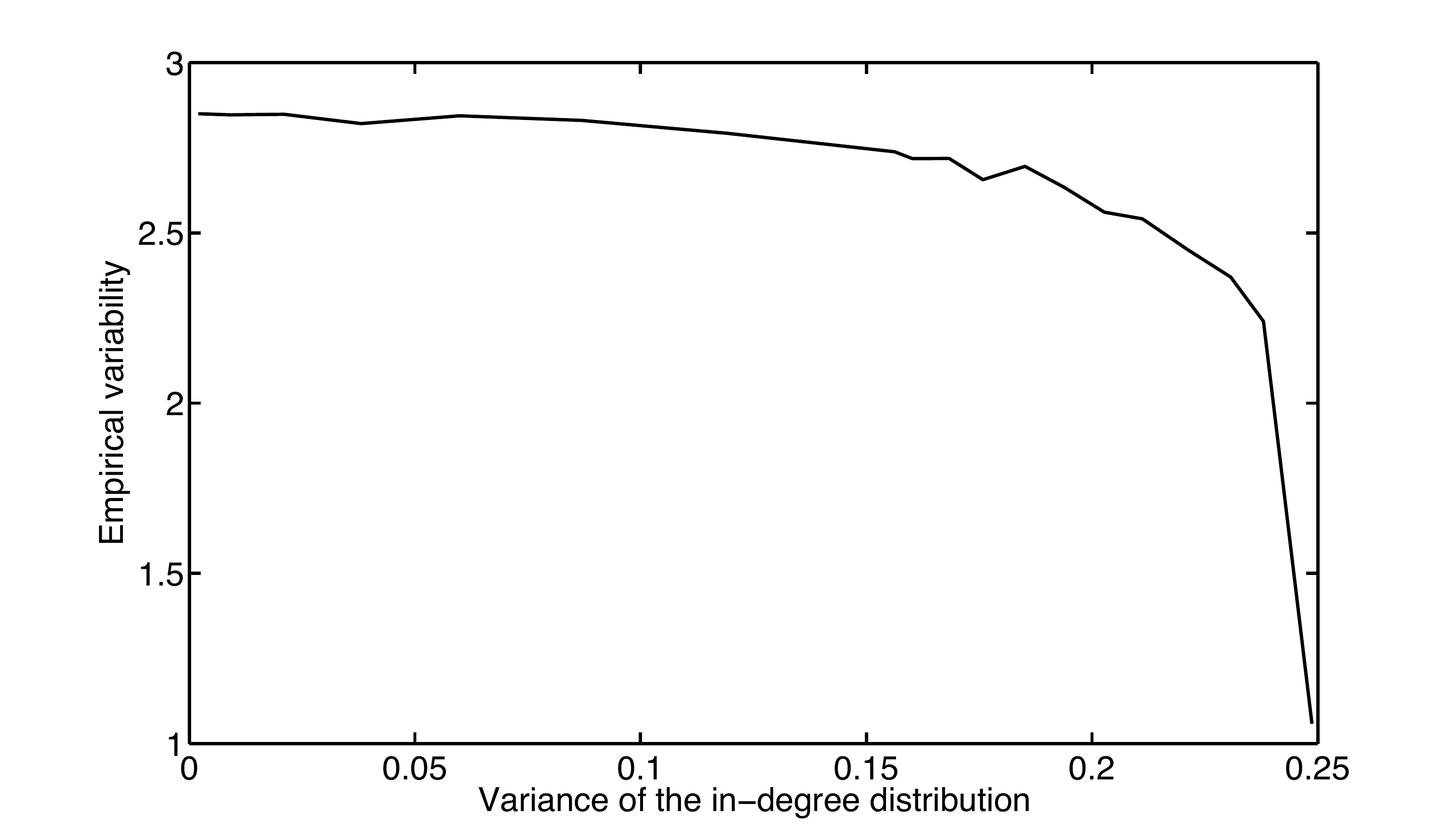}
	\caption{Estimated variability $\hat{\gamma}^2$ for the system of randomly coupled FitzHugh-Nagumo units, as a function of the in-degree variance. The family of in-degree distribution chosen here is $p=(\delta_c + \delta_{1-c})/2$, and the parameter $c$ is varied between $0$ and $0.5$ to generate networks with different variances $\mbox{Var}(\alpha)$ of the in-degree distribution. Parameters $n=500$, $\sigma=2$, $\alpha=1/3$, $\beta=0.08$, $\delta=0.05$, $\gamma=0.064$.}
\end{center}
\end{figure}

Therefore, in order to assess the generality of our results, we now consider a system of randomly connected Fitz-Hugh Nagumo (FHN) \cite{fitzhugh1961impulses,hermann2012heterogeneous} dynamical units. In contrast with system \eqref{eq:NN}, FHN units display the ability to generate action potentials, hence being closer to biological neurons behavior. More precisely, we consider the following system:
\begin{eqnarray}
	\dot{v}_i &=& v_i - \alpha v_i^3 - w_i + S\left(\sum_{j\to i} J_{ij}v_j\right)\\
	\dot{w}_i &=& \beta v - \gamma w +\delta 
\end{eqnarray}
where $(v_i,w_i)$ represents the state of neuron $i$, for $1\leq i \leq n$, and $\alpha,\beta,\gamma,\delta$ are real parameters. In Figure 2, we display the empirical variability of the variables $v_i$ as a function of the variance of the in-degree distribution, keeping the choice $p=(\delta_c + \delta_{1-c})/2$ for the sake of comparison, showing again that the variance is maximal when the network is regular. This numerical simulation supports the idea that, qualitatively, the phenomenon at stake does not depend much on the precise dynamical model, but is more a property of the underlying graph. Nevertheless, although decreasing, the shape of the curve displayed in Figure \ref{fig:2} is significantly different from the one displayed in Figure \ref{fig:1} : the impact of the in-degree distribution seems to be important only when the variance gets close to the maximal value $0.25$ corresponding to a network with half of the nodes having almost $n$ incoming inputs and the other half having almost zero incoming input. Understanding this specific profile would require an extension of the HMF theory to such non-linear interacting systems, for which the classical homogeneous mean-field theory still remains a complicated task.

Beyond the specific choice of the dynamical model for the neuron dynamics, a second limitation of our approach concerns the hypothesis made on the degree distribution. In particular, we assume that the degree distribution converges to the rescaled degree distribution in the large $n$ asymptotic. This assumption means that the typical number of incoming connections at each neuron is required to be of order $n$. In our setting, one cannot consider a network with a lot of neurons having very few incoming connections, for instance of order $1$. For those units, the number of incoming connections would be too small to apply, somehow, the central limit theorem which is at the heart, though hidden, of the mean-field theory. Understanding how this type of system would impact the HMF theory is a  subject of future research.

\section{Conclusion}

In this paper, we have shown that neural networks defined over regular graphs lead to a higher activity variability than those defined over irregular graphs. This somewhat counter-intuitive result illustrates that more homogeneity in the definition of a dynamical system does not necessarily lead to more order in the resulting dynamics. To obtain this result, we have developed upon the heterogeneous mean-field theory, providing general equations describing the statistical behavior of the neuronal population, showing a good agreement with numerical simulations. Moreover, we have also shown that the effect of increasing the variance of the network while increasing regularity does not interact strongly with the stability of the attractor whose largest Lyapunov exponent remains almost constant. Therefore, selecting the properties of the connectivity graph may provide a way to control the variability of a neural representation without influencing much its dynamical complexity or its sensitivity to small external perturbations.

\end{document}